\documentclass[preprint, prb, 12pt]{revtex4}
\usepackage{amsmath}
\usepackage{graphicx}
\usepackage{psfrag}
\usepackage{epsfig}

\def\be{\begin{equation}}
\def\ee{\end{equation}}
\def\bea{\begin{eqnarray}}
\def\eea{\end{eqnarray}}

\begin{document}


\title{Direct correlation function of square well fluid with wide well: First
order mean spherical approximation}

\author{S. P. Hlushak, A. Trokhymchuk}
\affiliation{Institute for Condensed Matter Physics, Svientsitskoho 1, 79011, Lviv,
Ukraine}
\author{S. Soko\l owski}
\affiliation{Department for the Modelling of Physico-Chemical Processes, Maria Curie
Sklodowska University, 20031 Lublin, Poland}

\begin{abstract}
An analytical expression for square-well fluid direct correlation function
(DCF) obtained recently by Tang (Y.Tang, J. Chem. Phys. \textbf{127}, 164504
(2007)) in the first-order mean spherical approximation is extended for
wider well widths ($2<\lambda < 3$). Theoretically obtained direct
correlation functions and radial distribution functions for square-well
fluid with $\lambda=2.1$ and $\lambda=2.5$ are compared with corresponding
results of Monte-Carlo simulation.

\end{abstract}

\date{}
\maketitle





\section{Introduction}

Recently, the systems of particles interacting with discrete potentials
gained much 
attention from the scientific community. Such an increased interest to this
class of model systems is associated primarely with their ability to mimic
the bulk properties of 
variety of complex fluid systems, like associating fluids, colloids, cluster
particles etc. The discrete hard-core potentials composed of an attractive
square well and a repulsive square barrier were shown to induce the
liquid-liquid phase transition in \ a single-component model fluid \cite%
{glaser2007,franzese2001gmg,rzysko2008jcp,cervantes2007jcp}. 
Similar phase transitions take place in the real substances like water,
carbon and phosphorus. Moreover, the effective potentials with square-well
plus square-barrier components may be used to mimic colloidal particles
interactions, which according to the DLVO theory are characterized by a
long-range repulsive barrier and one or possibly two attractive wells.
Finally, inhomogeneous fluids with descrete repulsive and attractive
interaction potentials represent another intriquing area of research because
they often serve as a benchmark to study a variety of interesting problems
such as interfacial phenomena, surface adsorption, wetting, capillary
condensation, etc.

The most suitable method to handle these systems seems to be the classical
density-functional theoriy (DFT). There are several versions of the DFT
derived usualy from either weighted density approximation or perturbation
expansions. Recently, Tang and Wu \cite{tang2003a1,tang2004a2,tang2004a3} 
have proposed a new DFT in which fundamental measure theory is combined with
the so-called first-order mean-spherical approximation (FMSA). Its
implementation requires knowledge of the thermodynamics and the direct
correlation function of \ the model bulk fluid. In comparison with
alternative approaches, DFT/FMSA possesses a number of advantages due to the
both accuracy and simplicity of the FMSA solution. In combination with the
DFT, the FMSA was already applied to study Lennard-Jones \cite{tang2003a1}, 
%
Yukawa \cite{tang2003a2}, 
%
Sutherlad \cite{Mi2008a4} 
%
fluid systems. Regarding the systems with descrete potentials, quite
recently \cite{tang2008a5} 
the DFT/FMSA approach has been applied to study the structural properties of
the square-well (SW) fluid model defined by the pair interaction potential,

\begin{equation}
u\left( r\right) =\left\{ 
\begin{array}{ll}
\infty , & \quad r<1, \\ 
-\varepsilon , & \quad 1\leq r<\lambda , \\ 
0, & \quad r\geq \lambda .%
\end{array}%
\right.  \label{SW}
\end{equation}%
with a unit hard-core diameter, the attractive strength $\varepsilon $ and
the well width parameter $1<\lambda <2$; the FMSA solution for the direct
correlation function (DCF) of this model system has been obtained in a
separate study \cite{tang2007dcf}. 
It has been shown that DFT/FMSA gives good performance for entire considered
range of attraction parameter $1<\lambda <2$.

\bigskip

The fact that FMSA is linear with respect to the interaction potential
allows one to use linear combinations of the different potential functions
for which FMSA solution is known. Regarding the case of the SW model this
means that one can employ the linear combinations of the square wells ($%
\varepsilon >0$) and/or square barriers ($\varepsilon <0$) in order to form
all variety of \ the discrete potential models. Unfortunately, majority of
potential applications of the discrete SW-based potential models (e.g., see
Refs. 
\cite{glaser2007,franzese2001gmg,rzysko2008jcp,cervantes2007jcp}) $\ \ $have
the range of interaction that exceeds one particle diameter, that requires
to consider $\lambda >2$ in Eq.(1). This means that existing FMSA solution
obtained by Tang \cite{tang2007dcf} is not sufficient to have FMSA be
applied in the studies of the complex discrete potential models. \ The aim
of present work is to extend the recent FMSA solution\cite{tang2007dcf} for
the SW model fluid with $1<\lambda <2$  to the case of the SW fluids with $%
2<\lambda <3$. Obtaining of a such FMSA solution for the attractive SW model
is crucial for the future application of both the FMSA 
as well as the DFT/FMSA to the systems with a combined (attractive SW plus
repulsive SB) discrete potential. This will be shown in a forthcoming paper
[]. The actual paper is organized as follows. In the next Section II we
outline the necessary details of the FMSA solution for SW potential with $%
1<\lambda <3.$\ The FMSA results for the direct and pair correlation
function will be presented and compared against computer simulation data in
Section III. We conclude the paper in Section IV.

\bigskip

\section{FMSA for square-well model fluids}

The general 
FMSA formalism, developed by Tang and Lu, is presented in their paper \cite%
{tang1993nso}. These authors solved the Ornstein-Zernike (OZ) equation, 
\begin{equation*}
\tilde{h}\left( k\right) =\tilde{c}\left( k\right) +\rho \tilde{h}\left(
k\right) \tilde{c}\left( k\right)
\end{equation*}%
for a one-component system of particles/moleculs of a number density $\rho $%
\ by employing the perturbative expansion for the total and direct
correlation functions, $h$ and $c,$ respectively, over the pair interaction
energy parameter $\varepsilon ,$ 
\begin{eqnarray}
\tilde{h}\left( k\right) &=&\tilde{h}_{0}\left( k\right) +\varepsilon \tilde{%
h}_{1}\left( k\right) +\dots  \label{h_pert} \\
\tilde{c}\left( k\right) &=&\tilde{c}_{0}\left( k\right) +\varepsilon \tilde{%
c}_{1}\left( k\right) +\dots ,  \label{c_pert}
\end{eqnarray}%
where the subscript $0$ denotes the contribution of a hard-sphere (HS)
reference system while the subscript $1$ stands for the first-order
perturbation term. Here and in what follows, in accordance with the previous
notation\cite{tang1993nso} all symbols with the tilde denote the
three-dimensional Fourier transforms, 
\begin{eqnarray}
\tilde{h}\left( k\right) &=&\frac{4\pi }{k}\int_{0}^{\infty }sin\left(
kr\right) rh\left( r\right) dr, \\
\tilde{c}\left( k\right) &=&\frac{4\pi }{k}\int_{0}^{\infty }sin\left(
kr\right) rc\left( r\right) dr,  \label{fourier3D}
\end{eqnarray}%
while all symbols with the hats denote the one-dimensional Fourier
transforms or the Laplace transforms.

By employing the Hilbert transform, the general solution of the first order
OZ equation can be obtained. 
The Fourier transform of the first order contribution to the total
correlation function $h$ reads, 
\begin{equation}
\hat{h}_{1}\left( k\right) =\frac{P\left( ik\right) }{{\hat{Q}_{0}^{2}\left(
ik\right) }},  \label{P_def}
\end{equation}%
where

\begin{equation}
P\left( ik\right) =\frac{U_{1}\left( k\right) }{2\hat{Q}_{0}^{2}\left(
-ik\right) }-\frac{e^{-ik{}}}{2i\pi }\int_{-\infty }^{\infty }\frac{%
U_{1}\left( y\right) e^{iy{}}}{\left( y-k\right) \hat{Q}_{0}^{2}\left(
-iy\right) }dy,  \label{P_k}
\end{equation}%
and $\hat{Q_{0}}\left( ik\right) $ is the Baxter hard-sphere factorization
function with a Laplace transform given by 
\begin{equation}
\hat{Q}_{0}\left( s\right) =\frac{S\left( s\right) +12\eta L\left( s\right)
e^{-s{}}}{\left( 1-\eta \right) ^{2}s^{3}},  \label{Q_0}
\end{equation}%
with $\eta =\frac{1}{6}\pi \rho $, and 
\begin{eqnarray}
S\left( t\right) &=&\left( 1-\eta \right) ^{2}t^{3}+6\eta \left( 1-\eta
\right) t^{2}+18\eta ^{2}t-12\eta \left( 1+2\eta \right) ,  \label{S_def} \\
L\left( t\right) &=&\left( 1+\frac{\eta }{2}\right) t+1+2\eta .
\label{L_def}
\end{eqnarray}%
The function $U_{1}\left( k\right) $ in Eq. (\ref{P_k}) is defined as

\begin{equation}
U_{1}\left( k\right) =\int_{1}^{\infty } rc_{1}\left( r\right) e^{-ikr}dr,
\label{U_1_def}
\end{equation}%
where, according to the FMSA closure, 
\begin{equation}
c_{1}\left( r\right) =-\beta u\left( r\right) ,\quad \mathrm{for}\quad r>1,
\label{c_1_def}
\end{equation}%
with $\beta =1/k_{B}T$ and $u\left( r\right) $ being the pair potential. 

For smooth pair potentials $u\left( r\right) $, for which $U_{1}\left(
k\right) \sim e^{-ik{}}$ (e.g., for the Yukava potential), the integration
contour in the right-hand side of Eq.~(\ref{P_k}) can be closed in the upper
complex half-plane and evaluation of the integral does not require to
calculate the residues at zeroes\ of the function $\hat{Q_{0}}\left(
-ik\right) $. This important simplification does not hold for the pair
potentials that vanish 
for $r>\lambda {}$ 
($\lambda >1$), e.g. for the SW potential. For such a potential Tang and Lu
expanded ${1}/{\hat{Q_{0}}\left( -ik\right) ^{2}}$ 
in a following way, 
\begin{equation}
\frac{1}{\hat{Q_{0}}\left( -ik\right) ^{2}}=\frac{\left( 1-\eta \right)
^{4}\left( -ik\right) ^{6}}{S^{2}\left( -ik\right) }\left[ 1-2\left( 12\eta 
\frac{L\left( -ik\right) }{S\left( -ik\right) }e^{ik}\right) +3\left( 12\eta 
\frac{L\left( -ik\right) }{S\left( -ik\right) }e^{ik}\right) ^{2}-\cdots %
\right] ,  \label{Q_expansion}
\end{equation}%
and performed calculations according to the scheme described in the Appendix
of their work\cite{tang1993nso}. Since only the first term of the expansion (%
\ref{Q_expansion}) survives for $\lambda \leq 2$, therefore 
\begin{equation}
\hat{h}_{1}\left( k\right) =-\frac{\left( 1-\eta \right) ^{4}e^{-ik}}{\hat{Q}%
_{0}^{2}\left( ik\right) }Res\left\{ \frac{W_{\lambda }\left( y\right)
\left( -iy\right) ^{6}e^{iy}}{\left( y-k\right) S^{2}\left( -iy\right) }%
\right\} ,\qquad \lambda \leq 2,  \label{h_1_Ll1}
\end{equation}%
where 
\begin{equation}
W_{\lambda }\left( k\right) =\int_{\lambda }^{\infty }r{c}_{1}\left(
r\right) e^{-ikr}dr  \label{W_lambda_def}
\end{equation}%
should be calculated for ${c}_{1}\left( r\right) $ smoothly extended from $%
r\in \left[ 1,\lambda \right) $ to $r\in \left[ \lambda ,\infty \right) $.
For the SW potential we get\cite{tang1994aas} 
\begin{equation}
W_{\lambda }\left( k\right) =\beta \varepsilon \frac{1+ik\lambda }{\left(
ik\right) ^{2}}e^{-ik\lambda }.  \label{W_lambda_SW_def}
\end{equation}

Expression (\ref{h_1_Ll1}) is valid only for the pair potentials $u\left(
r\right) $ with the attractive well width that does not exceed one particle
diameter, i.e., for $\lambda \leq 2$. In order to extend the actual FMSA
scheme for larger values of parameter $\lambda $, the contribution due to
the second term in the expansion (\ref{Q_expansion}) should be taken into
account. After doing that we obtain the following result, 
\begin{equation}
\hat{h}_{1}\left( k\right) =-\frac{\left( 1-\eta \right) ^{4}e^{-ik}}{\hat{Q}%
_{0}^{2}\left( ik\right) }Res\left\{ \frac{W_{\lambda }\left( y\right)
\left( -iy\right) ^{6}e^{iy}}{\left( y-k\right) S^{2}\left( -iy\right) }%
\left[ 1-24\eta \frac{L\left( -iy\right) }{S\left( -iy\right) }e^{iy}\right]
\right\} ,\quad 2<\lambda \leq 3,  \label{h_1_Ll2}
\end{equation}%
which may be viewed as Eq.~(\ref{h_1_Ll1}) supplemented with an additional
term, that contributes to $r>2$ region in the $r$-space.

The therrmodynamic and structural properties of the square-well fluid for $%
\lambda <2$ within the FMSA approximation 
have been reported in Refs.\cite{tang1994aas,tang2007dcf,tang1994for}. Here,
we would like to pay attention to the caase with $\lambda >2,$\ i.e., to the
development of an additional term that already appears in Eq.~(\ref{h_1_Ll2}%
). Let us denote it as 
\begin{equation}
\hat{h}_{1}^{\left( 2\right) }\left( k\right) =\frac{24\eta \left( 1-\eta
\right) ^{4}e^{-ik}}{\hat{Q}_{0}^{2}\left( ik\right) }Res\left\{ \frac{%
W_{\lambda }\left( y\right) \left( -iy\right) ^{6}L\left( -iy\right) e^{2iy}%
}{\left( y-k\right) S^{3}\left( -iy\right) }\right\} .  \label{h_1_L2}
\end{equation}%
After calculating the sum of residues in above expression, we obtain the
following Laplace transform 
\begin{eqnarray}
\hat{h}_{1}^{\left( 2\right) }\left( s\right) &=&\frac{24\eta \left( 1-\eta
\right) ^{4}e^{-s}}{\hat{Q}_{0}^{2}\left( s\right) }\left\{ -\frac{C\left(
-s\right) L\left( -s\right) e^{-s\left( \lambda -2\right) }}{S^{3}\left(
-s\right) }\right.  \notag \\
&+&\left. \frac{1}{2}\sum_{i=0}^{2}\left( \frac{G_{2}\left( t_{i}\right) }{%
s+t_{i}}+\frac{G_{1}\left( t_{i}\right) }{\left( s+t_{i}\right) ^{2}}+\frac{%
G_{0}\left( t_{i}\right) }{\left( s+t_{i}\right) ^{3}}\right) \right\} ,
\label{h_1_sL2}
\end{eqnarray}%
where $C\left( s\right) =\lambda s^{5}-s^{4}$, while $t_{i}$ correspond to
the roots of $S\left( t\right) =0$, and functions $G_{i}\left( s\right) $
are, 
\begin{eqnarray}
&&\!\!\!\!\!\!\!\!\!\!\!\!G_{0}\left( s\right) =-2\frac{C\left( s\right)
L\left( s\right) e^{s\left( \lambda -2\right) }}{S_{1}^{3}\left( s\right) },
\\
&&\!\!\!\!\!\!\!\!\!\!\!\!G_{1}\left( s\right) =2\frac{\left\{ \frac{{}}{{}}%
\frac{d\left[ C\left( s\right) L\left( s\right) \right] }{ds}+\left( \lambda
-2\right) C\left( s\right) L\left( s\right) \right\} e^{s\left( \lambda
-2\right) }}{S_{1}^{3}\left( s\right) }-3\frac{C\left( s\right) L\left(
s\right) S_{2}\left( s\right) e^{s\left( \lambda -2\right) }}{%
S_{1}^{4}\left( s\right) }, \\
&&\!\!\!\!\!\!\!\!\!\!\!\!G_{2}\left( s\right) =-\frac{\left\{ \frac{{}}{{}}%
\frac{d^{2}\left[ C\left( s\right) L\left( s\right) \right] }{ds^{2}}%
+2\left( \lambda -2\right) \frac{d\left[ C\left( s\right) L\left( s\right) %
\right] }{ds}+\left( \lambda -2\right) ^{2}C\left( s\right) L\left( s\right)
\right\} e^{s\left( \lambda -2\right) }}{S_{1}^{3}\left( s\right) } \\
&&+3\frac{\left\{ \frac{{}}{{}}\frac{d\left[ C\left( s\right) L\left(
s\right) \right] }{ds}+\left( \lambda -2\right) C\left( s\right) L\left(
s\right) \right\} S_{2}\left( s\right) e^{s\left( \lambda -2\right) }}{%
S_{1}^{4}\left( s\right) }+\frac{C\left( s\right) L\left( s\right)
S_{3}\left( s\right) e^{s\left( \lambda -2\right) }}{S_{1}^{4}\left(
s\right) }  \notag \\
&&-3\frac{C\left( s\right) L\left( s\right) S_{2}^{2}\left( s\right)
e^{s\left( \lambda -2\right) }}{S_{1}^{5}\left( s\right) },  \notag
\label{G2}
\end{eqnarray}%
with $S_{n}\left( s\right) =d^{n}S\left( s\right) /ds^{n}$.

Then the Laplace transform of the first order contribution to the total
correlation function $h$\ in the case of $2<\lambda \leq 3$\ can be written
as, 
\begin{equation*}
\hat{h}_{1}\left( s\right) =\frac{\beta \varepsilon \left( 1-\eta \right)
^{2}}{\hat{Q}_{0}^{2}\left( s\right) }\left[ {\hat{p}^{\left( 1\right)
}\left( s\right) }{}+24\eta {\hat{p}^{\left( 2\right) }\left( s\right) }{}%
\right] .
\end{equation*}%
In the above 
we introduced two functions 
\begin{equation}
\hat{p}^{\left( 1\right) }\left( s\right) =e^{-s}Res\left[ \frac{C\left(
t\right) e^{\left( \lambda -1\right) t}}{\left( s+t\right) S^{2}\left(
t\right) }\right] ,  \label{p1s}
\end{equation}%
and 
\begin{equation}
\hat{p}^{\left( 2\right) }\left( s\right) =e^{-s}Res\left[ \frac{C\left(
t\right) L\left( s\right) e^{\left( \lambda -2\right) t}}{\left( s+t\right)
S^{3}\left( t\right) }\right] .  \label{p2s}
\end{equation}%
The first one 
was examined in more details in Ref.\cite{tang2007dcf}, whereas the second
function, 
in agreement with Eq.~(\ref{h_1_sL2}), reads 
\begin{eqnarray}
\hat{p}^{\left( 2\right) }\left( s\right) &=&e^{-s}\left\{ -\frac{C\left(
-s\right) L\left( -s\right) e^{-s\left( \lambda -2\right) }}{S^{3}\left(
-s\right) }\right.  \notag \\
&&+\left. \frac{1}{2}\sum_{i=0}^{2}\left( \frac{G_{2}\left( t_{i}\right) }{%
s+t_{i}}+\frac{G_{1}\left( t_{i}\right) }{\left( s+t_{i}\right) ^{2}}+\frac{%
G_{0}\left( t_{i}\right) }{\left( s+t_{i}\right) ^{3}}\right) \right\} .
\label{p2s_2}
\end{eqnarray}

Following by Tang and Lu\cite{tang2007dcf}, the Laplace transform of the DCF 
can be obtained from the relation 
\begin{equation}
\hat{c}\left( s\right) =\beta \varepsilon \left( 1-\eta \right) ^{4}\left\{ %
\left[ \hat{Q_{p}}\left( s\right) \right] ^{\left[ 0,\infty \right] }-\left[ 
\hat{Q_{p}}\left( -s\right) \right] ^{\left[ 0,\infty \right] }\right\} ,
\label{c1s}
\end{equation}%
where superscript $\left[ 0,\infty \right] $ denotes the part 
that is nonzero only in $\left[ 0,\infty \right] $ region of the $r$ space,
and 
\begin{equation}
\hat{Q_{p}}\left( s\right) =\hat{Q_{0}^{2}}\left( -s\right) \left[ \hat{p}%
^{\left( 1\right) }\left( s\right) +24\eta \hat{p}^{\left( 2\right) }\left(
s\right) \right] .  \label{Qps}
\end{equation}%
In order to invert $\hat{c}\left( s\right) $ 
to the $r-$space, we should ignore all $\left[ 0,\infty \right] $
superscripts in relation (\ref{c1s}) and invert it to $r-$space. Since we
are interested only in $r>0$ region, we assume that ${c}\left( r\right) =0$
for $r<0$. Note, that the contribution from the first term $\hat{p}^{\left(
1\right) }\left( s\right) $ in Eq.~(\ref{Qps}) was already considered in Ref.%
\cite{tang2007dcf}. Let us examine now the second term of Eq.~(\ref{Qps}), 
\begin{equation}
\hat{Q_{p}}^{(2)}\left( s\right) =24\eta \hat{Q_{0}^{2}}\left( -s\right) 
\hat{p}^{\left( 2\right) }\left( s\right) .  \label{Qp2s}
\end{equation}%
Taking into account that $\hat{Q_{0}}\left( s\right) $ is analytical
function in the whole complex plane, we may simplify the expression for $%
\hat{p}^{\left( 2\right) }\left( s\right) $ by subtracting some analytical
function. Indeed, the subtracted analytical function being multiplied by $%
\hat{Q_{0}^{2}}\left( -s\right) $ will result in a new analytical function,
that has no impact on the resulting expressions for $c_{1}\left( r\right) $
in the $r-$space. In such a way we replace the first term $-{C\left(
-s\right) L\left( -s\right) e^{-s\left( \lambda -2\right) }}/{S^{3}\left(
-s\right) }$ in Eq. (\ref{p2s_2}) by expression $\sum\limits_{i=0}^{2}\left( 
\frac{C_{3}\left( t_{i}\right) }{s+t_{i}}+\frac{C_{2}\left( t_{i}\right) }{%
\left( s+t_{i}\right) ^{2}}+\frac{C_{1}\left( t_{i}\right) }{\left(
s+t_{i}\right) ^{3}}\right) e^{-s\left( \lambda -2\right) }$, where the
coefficients $C_{1}\left( t_{i}\right) $, $C_{2}\left( t_{i}\right) $ and $%
C_{3}\left( t_{i}\right) $ are found from the equality of residues at all
poles $t_{i}$ of the expression 
\begin{equation}
\underset{\{t_{i}\}}{Res}\left[ \left( \frac{C_{3}\left( t_{i}\right) }{%
s+t_{i}}+\frac{C_{2}\left( t_{i}\right) }{\left( s+t_{i}\right) ^{2}}+\frac{%
C_{1}\left( t_{i}\right) }{\left( s+t_{i}\right) ^{3}}\right) e^{-s\left(
\lambda -2\right) }\right] =\underset{\{t_{i}\}}{Res}\left[ -\frac{C\left(
-s\right) L\left( -s\right) e^{-s\left( \lambda -2\right) }}{S^{3}\left(
-s\right) }\right] .  \label{res_equality}
\end{equation}%
The new simplified function $\hat{p}^{\prime \left( 2\right) }\left(
s\right) $ reads 
\begin{eqnarray}
\hat{p}^{\prime \left( 2\right) }\left( s\right)
&=&\sum\limits_{i=0}^{2}\left( \frac{C_{3}\left( t_{i}\right) }{s+t_{i}}+%
\frac{C_{2}\left( t_{i}\right) }{\left( s+t_{i}\right) ^{2}}+\frac{%
C_{1}\left( t_{i}\right) }{\left( s+t_{i}\right) ^{3}}\right) e^{-s\left(
\lambda -1\right) }  \notag \\
&&+\frac{1}{2}\sum_{i=0}^{2}\left( \frac{G_{2}\left( t_{i}\right) }{s+t_{i}}+%
\frac{G_{1}\left( t_{i}\right) }{\left( s+t_{i}\right) ^{2}}+\frac{%
G_{0}\left( t_{i}\right) }{\left( s+t_{i}\right) ^{3}}\right) e^{-s},
\label{ps_prime}
\end{eqnarray}%
where 
\begin{eqnarray}
C_{1}\left( t_{i}\right) &=&\frac{C\left( t_{i}\right) L\left( t_{i}\right) 
}{S_{1}^{3}\left( t_{i}\right) },  \notag \\
C_{2}\left( t_{i}\right) &=&-\frac{\frac{d}{ds}\left[ C\left( s\right)
L\left( s\right) \right] _{s=t_{i}}}{S_{1}^{3}\left( t_{i}\right) }+\frac{3}{%
2}\frac{C\left( t_{i}\right) L\left( t_{i}\right) S_{2}\left( t_{i}\right) }{%
S_{1}^{4}\left( t_{i}\right) },  \notag \\
C_{3}\left( t_{i}\right) &=&\frac{\frac{d^{2}}{ds^{2}}\left[ C\left(
s\right) L\left( s\right) \right] _{s=t_{i}}}{2S_{1}^{3}\left( t_{i}\right) }%
-\frac{3}{2}\frac{\frac{d}{ds}\left[ C\left( s\right) L\left( s\right) %
\right] _{s=t_{i}}S_{2}\left( t_{i}\right) }{S_{1}^{4}\left( t_{i}\right) } 
\notag \\
&&-\frac{C\left( t_{i}\right) L\left( t_{i}\right) S_{3}\left( t_{i}\right) 
}{2S_{1}^{4}\left( t_{i}\right) }+\frac{3}{2}\frac{C\left( t_{i}\right)
L\left( t_{i}\right) S_{2}^{2}\left( t_{i}\right) }{S_{1}^{5}\left(
t_{i}\right) }.  \notag  \label{C_i}
\end{eqnarray}

It is easy to find that the functions $G_{i}\left( s\right) $ from Eqs.
(21)-(23) can be expressed in terms of $C_{i}\left( s\right) $ in the
following way 
\begin{eqnarray}
G_{0}\left( s\right) &=&-2C_{1}\left( s\right) e^{\left( \lambda -2\right)
s},  \notag \\
G_{1}\left( s\right) &=&-2\left[ C_{2}\left( s\right) -\left( \lambda
-2\right) C_{1}\left( s\right) \right] e^{\left( \lambda -2\right) s}, 
\notag \\
G_{2}\left( s\right) &=&-2\left[ C_{3}\left( s\right) -\left( \lambda
-2\right) C_{2}\left( s\right) +\frac{\left( \lambda -2\right) ^{2}}{2}%
C_{1}\left( s\right) \right] e^{\left( \lambda -2\right) s}.  \notag
\label{GviaC}
\end{eqnarray}%
Then the inverse Laplace transform of Eq.~(\ref{ps_prime}) reads 
\begin{eqnarray}
{p}^{\prime \left( 2\right) }\left( r\right) &=&\frac{1}{2}%
\sum_{i=0}^{2}\left( \underset{\frac{{}}{{}}}{}G_{2}\left( t_{i}\right)
+\left( r-1\right) G_{1}\left( t_{i}\right) \right.  \notag \\
&&+\left. \frac{\left( r-1\right) ^{2}}{2}G_{0}\left( t_{i}\right) \right)
e^{-t_{i}\left( r-1\right) }\left[ H\left( r-1\right) -H\left( r-\lambda
+1\right) \right] ,  \label{p2r_prime}
\end{eqnarray}%
which is nonzero inside $\left[ 1,\lambda -1\right] $, and $H\left( r\right)$ 
is the Heaviside step function. Substituting (\ref{ps_prime}) and 
(\ref{Q_0}) into (\ref{Qp2s}), we obtain 
\begin{eqnarray}
&&\frac{\hat{Q_{p}}^{(2)}\left( s\right) }{24\eta }=\hat{p}^{\prime \left(
2\right) }\left( s\right) +\left[ \sum_{i=1}^{6}\frac{e_{0,i}}{\left(
-s\right) ^{i}}+\sum_{i=2}^{6}\frac{e_{1,i}}{\left( -s\right) ^{i}}%
e^{s}+\sum_{i=4}^{6}\frac{e_{2,i}}{\left( -s\right) ^{i}}e^{2s}\right]
\sum_{i=0}^{2}\left\{ \frac{C_{3}\left( t_{i}\right) }{s+t_{i}}\right. 
\notag \\
&&+\left. \frac{C_{2}\left( t_{i}\right) }{\left( s+t_{i}\right) ^{2}}+\frac{%
C_{1}\left( t_{i}\right) }{\left( s+t_{i}\right) ^{3}}\right\} e^{-\left(
\lambda -1\right) s}+\frac{1}{2}\left[ \sum_{i=1}^{6}\frac{e_{0,i}}{\left(
-s\right) ^{i}}+\sum_{i=2}^{6}\frac{e_{1,i}}{\left( -s\right) ^{i}}%
e^{s}\right.  \notag \\
&&+\left. \sum_{i=4}^{6}\frac{e_{2,i}}{\left( -s\right) ^{i}}e^{2s}\right]
\sum_{i=0}^{2}\left\{ \frac{G_{2}\left( t_{i}\right) }{s+t_{i}}+\frac{%
G_{1}\left( t_{i}\right) }{\left( s+t_{i}\right) ^{2}}+\frac{G_{0}\left(
t_{i}\right) }{\left( s+t_{i}\right) ^{3}}\right\} e^{-s},
\label{Qp2s_prime}
\end{eqnarray}%
and 
\begin{eqnarray}
&&\frac{\hat{Q_{p}}^{(2)}\left( -s\right) }{24\eta }=\hat{p}^{\prime \left(
2\right) }\left( -s\right) -\left[ \sum_{i=1}^{6}\frac{e_{0,i}}{s^{i}}%
+\sum_{i=2}^{6}\frac{e_{1,i}}{s^{i}}e^{-s}+\sum_{i=4}^{6}\frac{e_{2,i}}{s^{i}%
}e^{-2s}\right] \sum_{i=0}^{2}\left\{ \frac{C_{3}\left( t_{i}\right) }{%
s-t_{i}}\right.  \notag \\
&&-\left. \frac{C_{2}\left( t_{i}\right) }{\left( s-t_{i}\right) ^{2}}+\frac{%
C_{1}\left( t_{i}\right) }{\left( s-t_{i}\right) ^{3}}\right\} e^{\left(
\lambda -1\right) s}-\frac{1}{2}\left[ \sum_{i=1}^{6}\frac{e_{0,i}}{s^{i}}%
+\sum_{i=2}^{6}\frac{e_{1,i}}{s^{i}}e^{-s}\right.  \notag \\
&&+\left. \sum_{i=4}^{6}\frac{e_{2,i}}{s^{i}}e^{-2s}\right]
\sum_{i=0}^{2}\left\{ \frac{G_{2}\left( t_{i}\right) }{s-t_{i}}-\frac{%
G_{1}\left( t_{i}\right) }{\left( s-t_{i}\right) ^{2}}+\frac{G_{0}\left(
t_{i}\right) }{\left( s-t_{i}\right) ^{3}}\right\} e^{s},
\label{Qp2ms_prime}
\end{eqnarray}%
where the coefficients $e_{n,i}$ of the expansion\ of the Baxter function $%
\hat{Q_{0}}^{2}\left( s\right) $ are given in the Appendix.

In order to obtain the DCF 
in $r-$space we will use the method that is similar to that outlined in Ref.%
\cite{tang2007dcf}. Namely, from Eqs.~(\ref{Qp2s_prime}) and (\ref%
{Qp2ms_prime}) it is evident, that the key expressions to be inverted are
the functions of following form 
\begin{equation}
\hat{w}\left( s,z,n,m\right) =\frac{1}{s^{n}\left( s+z\right) ^{m}}=\left\{ 
\begin{array}{ll}
\sum\limits_{k=1}^{n}\frac{a_{k}^{nm}}{s^{k}}+\sum\limits_{k=1}^{m}\frac{%
b_{k}^{nm}}{\left( s+z\right) ^{k}}, & z\neq 0, \\ 
\frac{1}{s^{m+n}}, & z=0,%
\end{array}%
\right.  \label{wsznm}
\end{equation}%
with coefficients 
\begin{eqnarray}
a_{k}^{nm} &=&\left( -1\right) ^{n-m}\frac{\left( m+n-k-1\right) !}{\left(
n-k\right) !\left( m-1\right) !z^{m+n-k}}, \\
b_{k}^{nm} &=&\left( -1\right) ^{n}\frac{\left( m+n-k-1\right) !}{\left(
m-k\right) !\left( n-1\right) !z^{m+n-k}}.  \label{abnmk}
\end{eqnarray}%
%
%
%
%
%
%
%
%
%
The inverse Laplace transform of these 
functions reads 
\begin{equation}
\hat{w}\left( r,z,n,m\right) =\left\{ 
\begin{array}{ll}
\left[ \sum\limits_{k=1}^{n}a_{k}^{nm}\frac{r^{k-1}}{\left( k-1\right) !}%
+\sum\limits_{k=1}^{m}b_{k}^{nm}\frac{r^{k-1}}{\left( k-1\right) !}e^{-zr}%
\right] H\left( r\right) , & z\neq 0, \\ 
\frac{r^{m+n-1}}{\left( m+n-1\right) !}H\left( r\right) , & z=0,%
\end{array}%
\right.  \label{wrznm}
\end{equation}%
where $H\left( r\right) $ is the Heaviside step function. With this in
hands, we may now write down in terms of $\hat{w}\left( r,z,n,m\right) $\
the expressions for the inverse Laplace transforms of Eqs.~(\ref{Qp2s_prime}%
) and (\ref{Qp2ms_prime}) for $r\geq 0,$ 
\begin{eqnarray}
&&{\!\!\!\!\!\!\!\!\!\!\!\!\!\!}\frac{Q_{p}^{\left( 2\right) }\left(
r\right) }{24\eta }=p^{\left( 2\right) }\left( r\right)
+\sum_{i=1}^{6}\sum_{j=0}^{2}\frac{e_{0,i}}{\left( -1\right) ^{i}}\left[ 
\frac{{}}{{}}C_{3}\left( t_{j}\right) w\left( r-\lambda +1,t_{j},i,1\right)
\right.  \notag \\
&&\left. +C_{2}\left( t_{j}\right) w\left( r-\lambda +1,t_{j},i,2\right)
+C_{1}\left( t_{j}\right) w\left( r-\lambda +1,t_{j},i,3\right) \frac{{}}{{}}%
\right] H\left( r-\lambda +1\right)  \notag \\
&&{\!\!\!\!\!\!\!\!\!\!\!\!\!\!}+\sum_{i=2}^{6}\sum_{j=0}^{2}\frac{e_{1,i}}{%
\left( -1\right) ^{i}}\left[ \frac{{}}{{}}C_{3}\left( t_{j}\right) w\left(
r-\lambda +2,t_{j},i,1\right) +C_{2}\left( t_{j}\right) w\left( r-\lambda
+2,t_{j},i,2\right) \right.  \notag \\
&&\left. +C_{1}\left( t_{j}\right) w\left( r-\lambda +2,t_{j},i,3\right) 
\frac{{}}{{}}\right] H\left( r-\lambda +2\right)  \notag \\
&&{\!\!\!\!\!\!\!\!\!\!\!\!\!\!}+\sum_{i=4}^{6}\sum_{j=0}^{2}\frac{e_{2,i}}{%
\left( -1\right) ^{i}}\left[ \frac{{}}{{}}C_{3}\left( t_{j}\right) w\left(
r-\lambda +3,t_{j},i,1\right) \right.  \notag \\
&&\left. +C_{2}\left( t_{j}\right) w\left( r-\lambda +3,t_{j},i,2\right)
+C_{1}\left( t_{j}\right) w\left( r-\lambda +3,t_{j},i,3\right) \frac{{}}{{}}%
\right] H\left( r-\lambda +3\right)  \notag \\
&&{\!\!\!\!\!\!\!\!\!\!\!\!\!\!}+\frac{1}{2}\sum_{i=1}^{6}\sum_{j=0}^{2}%
\frac{e_{0,i}}{\left( -1\right) ^{i}}\left[ \frac{{}}{{}}G_{2}\left(
t_{j}\right) w\left( r-1,t_{j},i,1\right) \right.  \notag \\
&&\left. +G_{1}\left( t_{j}\right) w\left( r-1,t_{j},i,2\right) +G_{0}\left(
t_{j}\right) w\left( r-1,t_{j},i,3\right) \frac{{}}{{}}\right] H\left(
r-1\right)  \notag \\
&&{\!\!\!\!\!\!\!\!\!\!\!\!\!\!}+\frac{1}{2}\sum_{i=2}^{6}\sum_{j=0}^{2}%
\frac{e_{1,i}}{\left( -1\right) ^{i}}\left[ \frac{{}}{{}}G_{2}\left(
t_{j}\right) w\left( r,t_{j},i,1\right) \right.  \notag \\
&&\left. +G_{1}\left( t_{j}\right) w\left( r,t_{j},i,2\right) +G_{0}\left(
t_{j}\right) w\left( r,t_{j},i,3\right) \frac{{}}{{}}\right] H\left( r\right)
\notag \\
&&{\!\!\!\!\!\!\!\!\!\!\!\!\!\!}+\frac{1}{2}\sum_{i=4}^{6}\sum_{j=0}^{2}%
\frac{e_{2,i}}{\left( -1\right) ^{i}}\left[ \frac{{}}{{}}G_{2}\left(
t_{j}\right) w\left( r+1,t_{j},i,1\right) \right.  \notag \\
&&\left. +G_{1}\left( t_{j}\right) w\left( r+1,t_{j},i,2\right) +G_{0}\left(
t_{j}\right) w\left( r+1,t_{j},i,3\right) \frac{{}}{{}}\right] H\left(
r+1\right) ,  \label{Q2pr}
\end{eqnarray}%
\begin{eqnarray}
&&{\!\!\!\!\!\!\!\!\!\!\!\!\!\!}\frac{Q_{p}^{\left( 2\right) }\left(
-r\right) }{24\eta }=p^{\left( 2\right) }\left( -r\right)
-\sum_{i=1}^{6}\sum_{j=0}^{2}e_{0,i}\left[ \frac{{}}{{}}C_{3}\left(
t_{j}\right) w\left( r+\lambda -1,-t_{j},i,1\right) \right.  \notag \\
&&\left. -C_{2}\left( t_{j}\right) w\left( r+\lambda -1,-t_{j},i,2\right)
+C_{1}\left( t_{j}\right) w\left( r+\lambda -1,-t_{j},i,3\right) \frac{{}}{{}%
}\right] H\left( r+\lambda -1\right)  \notag \\
&&{\!\!\!\!\!\!\!\!\!\!\!\!\!\!}-\sum_{i=2}^{6}\sum_{j=0}^{2}e_{1,i}\left[ 
\frac{{}}{{}}C_{3}\left( t_{j}\right) w\left( r+\lambda -2,-t_{j},i,1\right)
\right.  \notag \\
&&\left. -C_{2}\left( t_{j}\right) w\left( r+\lambda -2,-t_{j},i,2\right)
+C_{1}\left( t_{j}\right) w\left( r+\lambda -2,-t_{j},i,3\right) \frac{{}}{{}%
}\right] H\left( r+\lambda -2\right)  \notag \\
&&{\!\!\!\!\!\!\!\!\!\!\!\!\!\!}-\sum_{i=4}^{6}\sum_{j=0}^{2}e_{2,i}\left[ 
\frac{{}}{{}}C_{3}\left( t_{j}\right) w\left( r+\lambda -3,-t_{j},i,1\right)
\right.  \notag \\
&&\left. -C_{2}\left( t_{j}\right) w\left( r+\lambda -3,-t_{j},i,2\right)
+C_{1}\left( t_{j}\right) w\left( r+\lambda -3,-t_{j},i,3\right) \frac{{}}{{}%
}\right] H\left( r+\lambda -3\right)  \notag \\
&&{\!\!\!\!\!\!\!\!\!\!\!\!\!\!}-\frac{1}{2}\sum_{i=1}^{6}%
\sum_{j=0}^{2}e_{0,i}\left[ \frac{{}}{{}}G_{2}\left( t_{j}\right) w\left(
r+1,-t_{j},i,1\right) \right.  \notag \\
&&\left. -G_{1}\left( t_{j}\right) w\left( r+1,-t_{j},i,2\right)
+G_{0}\left( t_{j}\right) w\left( r+1,-t_{j},i,3\right) \frac{{}}{{}}\right]
H\left( r+1\right)  \notag \\
&&{\!\!\!\!\!\!\!\!\!\!\!\!\!\!}-\frac{1}{2}\sum_{i=2}^{6}%
\sum_{j=0}^{2}e_{1,i}\left[ \frac{{}}{{}}G_{2}\left( t_{j}\right) w\left(
r,-t_{j},i,1\right) \right.  \notag \\
&&\left. -G_{1}\left( t_{j}\right) w\left( r,-t_{j},i,2\right) +G_{0}\left(
t_{j}\right) w\left( r,-t_{j},i,3\right) \frac{{}}{{}}\right] H\left(
r\right)  \notag \\
&&{\!\!\!\!\!\!\!\!\!\!\!\!\!\!}-\frac{1}{2}\sum_{i=4}^{6}%
\sum_{j=0}^{2}e_{2,i}\left[ \frac{{}}{{}}G_{2}\left( t_{j}\right) w\left(
r-1,-t_{j},i,1\right) \right.  \notag \\
&&\left. -G_{1}\left( t_{j}\right) w\left( r-1,-t_{j},i,2\right)
+G_{0}\left( t_{j}\right) w\left( r-1,-t_{j},i,3\right) \frac{{}}{{}}\right]
H\left( r-1\right) .  \label{Q2pmr}
\end{eqnarray}%
%
%
%
%
%
%
%
%
%
The above expressions are then substituted into inverse of Eq.~(\ref{Qps}), 
\begin{equation}
Q_{p}\left( r\right) =Q_{p}^{\left( 1\right) }\left( r\right) +Q_{p}^{\left(
2\right) }\left( r\right) ,  \label{Qpr}
\end{equation}%
where $Q_{p}^{\left( 1\right) }\left( r\right) $ is the inverese Laplace
transform of the function $\hat{Q_{p}}^{(1)}\left( s\right) =24\eta \hat{%
Q_{0}^{2}}\left( -s\right) \hat{p}^{\left( 1\right) }\left( s\right) $\ and
describes the contribution of the first term of the expansion (\ref{h_1_Ll2}%
); the expression for this contribution was given 
in Ref.\cite{tang2007dcf}. Then the first-order perturbation contribution to
the DCF 
of the SW model fluid with the radius of interaction that exceeds two
particle diameters reads, 
\begin{equation*}
rc_{1}\left( r\right) =\beta \varepsilon \left( 1-\eta \right) ^{4}\left[
Q_{p}\left( r\right) -Q_{p}\left( -r\right) \right] .
\end{equation*}%
The equations (\ref{Q2pr}) and (\ref{Q2pmr}) have been presented here in a
similar manner as it has been done in the work of Tang and Lu\cite%
{tang2007dcf} for the case of the SW model fluid with the range of
interaction that does not exceed two particle diameters, i.e. for $\ \lambda
\leq 2.$ 
The resulting expression for $c_{1}\left( r\right) $ in the case of $\lambda
>2$\ is consequently 
rather long, but also quite straightforward. It may be summarized as
follows, 
\begin{equation}
rc_{1}\left( r\right) =\left\{ 
\begin{array}{ll}
0, & r>\lambda , \\ 
\beta \varepsilon r, & 1<r\leq \lambda , \\ 
\beta \varepsilon \left( 1-\eta \right) ^{4}\left[ Q_{p}\left( r\right)
-Q_{p}\left( -r\right) \right] , & r\leq 1,%
\end{array}%
\right.  \label{rc1f}
\end{equation}%
where $Q_{p}\left( r\right) $ is given by Eq.~(\ref{Qpr}). The direct
correlation function is discontinuous at $r=1$ and $r=\lambda $.\ \ The
continuity of the DCF 
inside the core, $r\in \left[ 0,1\right] $, is evident from Eqs.~(\ref%
{Qp2s_prime}) and (\ref{Qp2ms_prime}), where the terms, that might
contribute to the discontinuities at $r=\lambda -2$ and $r=3-\lambda $, are
at least of the second order in ${1}/{s}$.


\bigskip

\section{Results and discussions}

To examine the evolution experienced by the DCF of the SW fluid upon
increase of the attractive well width, in Fig.~\ref{fig:DCF_SWr8} we show
the first-order DCF $c_{1}\left( r\right) $ 
evaluated for nine different values of width parameter $\lambda $ \ in the
range from $\lambda =1.5$ to $\lambda =3.0.$\ \ All calculations are
performed for fixed density $\rho =0.75$ and temperature $\beta \varepsilon
=0.5$. Note, that  the dependence of $c_{1}\left( r\right) $ on the
attractive well width $\lambda $ \ in the core region \ $r<1$\ \ is not
trivial. 

It is very well illustrated by the behavior of the value of $c_{1}\left(
r=0\right).$ One can see that value of $c_{1}\left( r=0\right)$ is first
decreasing upon $\lambda $ increases from 
$\lambda =1.5$ to $\lambda \approx 1.7$ and after this starts to grow with 
$\lambda $ increases. However, after reaching the value of $\lambda \approx
2.2$, $c_{1}\left( r=0\right)$ starts to drop till the value of
$\lambda \approx 2.6$ after which it begins to grow again.


In Figures~\ref{fig:dcf_rdf_l25}~and~\ref{fig:dcf_rdf_l21} we show the total
DCF $c\left( r\right) $ and total radial distribution function (RDF) $%
g\left( r\right) =h_{1}\left( r\right) +1$ of the SW fluid with attractive
well width $\lambda =2.1$ and $2.5$\ \ \ at the density $\rho =0.75.$ Both
functions are obtained within the FMSA and from Monte-Carlo simulations. The
FMSA results for the first-order correction term $h_{1}\left( r\right) $ to
the RDF were obtained numerically. The correspondin DCF and RDF of the
hard-sphere reference system, $c_{0}\left( r\right) $\ and $g_{0}\left(
r\right) $, respectively, were obtained from the first-order GMSA theory of
Tang and Lu \cite{tang1995ier}. The results for the intermediate and low
densities are not shown, since at such conditions the considered SW fluid
seems to be unstable. Good agreement between Monte Carlo simulation data and
\ FMSA theory results could be observed for the SW fluid with $\lambda =2.5$
(see Fig.~\ref{fig:dcf_rdf_l25}). The same doesn't hold for the SW fluid
with $\lambda =2.1$. The discrepancies between FMSA and simulation
predictions are clearly visible for both DCF and  RDF at the particle
contact $r=1$ and at the point of discontinuity at $r=\lambda =2.1$. These
shortcomings in the case of RDF can be corrected by employing exponential
(EXP) \cite{chandler1972oce} 
\begin{equation}
g^{EXP}\left( r\right) =g_{0}\left( r\right) e^{h_{1}\left( r\right) }
\label{EXP_appr}
\end{equation}%
or linearized exponential (LEXP) \cite{verlet1974ptp} 
\begin{equation}
g^{LEXP}\left( r\right) =g_{0}\left( r\right) \left( 1+h_{1}\left( r\right)
\right)   \label{LEXP_appr}
\end{equation}%
approximations. Indeed, the LEXP approximation significantly improves the
RDF of the SW fluid, giving better contact value at $r=1$ and improving RDF
values on both sides of the discontinuity at $r=\lambda =2.1$ (see Fig.~ \ref%
{fig:dcf_rdf_l21}, right panel).

\bigskip

In Fig.~\ref{fig:Compress} we present the inverse reduced isothermal
compressibilities of the SW fluids that was obtained from

\begin{equation}
\frac{1}{\chi _{T}}=\beta \left( \frac{\partial P}{\partial \rho }\right)
_{T}=1-4\pi \rho \int c\left( r\right) r^{2}dr,  \label{fluct_eq}
\end{equation}

\bigskip

The DCFs of the reference system of hard spheres were obtained from the
first-order GMSA theory of Tang and Lu \cite{tang1995ier}. From Fig.~\ref%
{fig:Compress} we note that at the high densities ($\eta\approx 0.5$) the
compressibility of the SW fluid with $\lambda=2.5$ increases rapidly and
almost reaches the level of the fluid $\lambda=2.0$. Mathematically, this
can be explained by positive contribution into compressibility of the
first-order DCF in the $r<1$ region, which compensates the negative
contribution due to the potential well. 

On Fig.\ref{fig:spinodal} we present spinodal curves and critical points 
for few long-range SW fluids obtained from condition
$\left( \frac{\partial P}{\partial \rho}\right)_{T}=0 $ and (\ref{fluct_eq}).

\section{Conclusions}

We extended the FMSA theory to deal with the square-well fluids of large
wells width ($1<\lambda <3$) than those studied previously and obtained
analytical expressions for the DCF. We obtained reasonable agreement of the
DCF and RDFs with MC simulation data and proposed exponential (EXP) and
linearized exponential (LEXP) approximations to correct RDF values close to
the contact.

\section*{Acknowledgments}

AT thanks MCSU for the hospitality while visiting the Department for the
Modelling of Physico-Chemical Processes when this project has been initiated.

\appendix

\section{Expansions of the Laplace transform of the Baxter factorization
function}

Laplace transforms of the Baxter hard-sphere factorization function and
square of the transform are 
\begin{equation}
\hat{Q_0}\left(s\right)=1+\sum_{i=1}^3\frac{a_i}{s_i} +\sum_{i=2}^3\frac{b_i%
}{s_i}e^{-s},  \label{Q_0app}
\end{equation}
and 
\begin{equation}
\hat{Q_0}^2\left(s\right)=1+\sum_{i=1}^6\frac{e_{0,i}}{s^i} +\sum_{i=2}^6%
\frac{e_{1,i}}{s^i}e^{-s} +\sum_{i=4}^6\frac{e_{2,i}}{s^i}e^{-2s},
\label{Q0s2app}
\end{equation}
where 
\begin{eqnarray}
&& b_2=\frac{12\eta\left( 1+\frac{\eta}{2} \right)}{\left( 1-\eta \right)^2}%
, \quad b_3=\frac{12\eta\left(1+{2\eta}\right)}{\left( 1-\eta \right)^2}, 
\notag \\
&& a_1=b_2-\frac{b_3}{2},\quad a_2=b_3-b_2,\quad a_3=-b_3,  \label{ai_bi_app}
\end{eqnarray}
and 
\begin{eqnarray}
&& e_{0,1}=2a_1,\quad e_{0,2}=a_1^2+2a_2, \quad e_{0,3}=2a_3+2a_1a_2,  \notag
\\
&& e_{0,4}=2a_1a_3+a_2^2,\quad e_{0,5}=2a_2a_3,\quad e_{0,6}=a_3^2,  \notag
\\
&& e_{1,2}=2b_2,\quad e_{1,3}=2b_3+2a_1b_2,\quad e_{1,4}=2a_1b_3+2a_2b_2, 
\notag \\
&& e_{1,5}=2a_2b_3+2a_3b_2,\quad e_{1,6}=2a_3b_3,  \notag \\
&& e_{2,4}=b_2^2,\quad e_{2,5}=2b_2b_3,\quad e_{2,6}=b_3^2.  \notag
\label{e_ni_app}
\end{eqnarray}



\newpage

\centerline{FIGURE CAPTIONS}

\begin{figure}[tbp]
\includegraphics[width=12cm, clip]{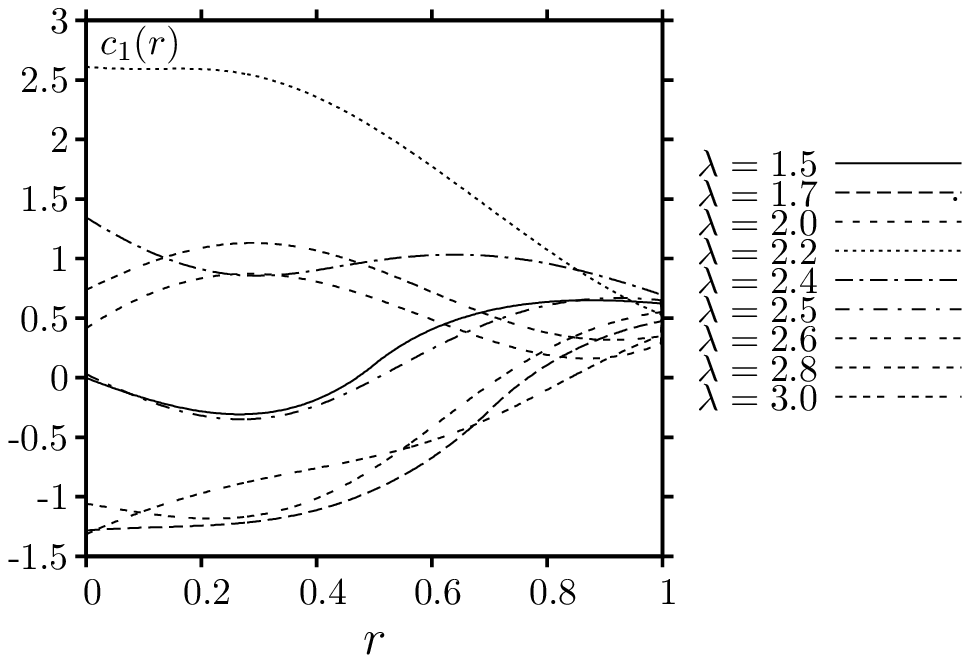} 
\caption{}
\label{fig:DCF_SWr8}
\end{figure}

\textbf{Fig.~\ref{fig:DCF_SWr8} First-order DCF for several $\lambda$ values
with $\varepsilon\beta=0.5$ and $\rho=0.75$. }

\begin{figure}[tbp]
\begin{center}
\includegraphics[width=12cm, clip]{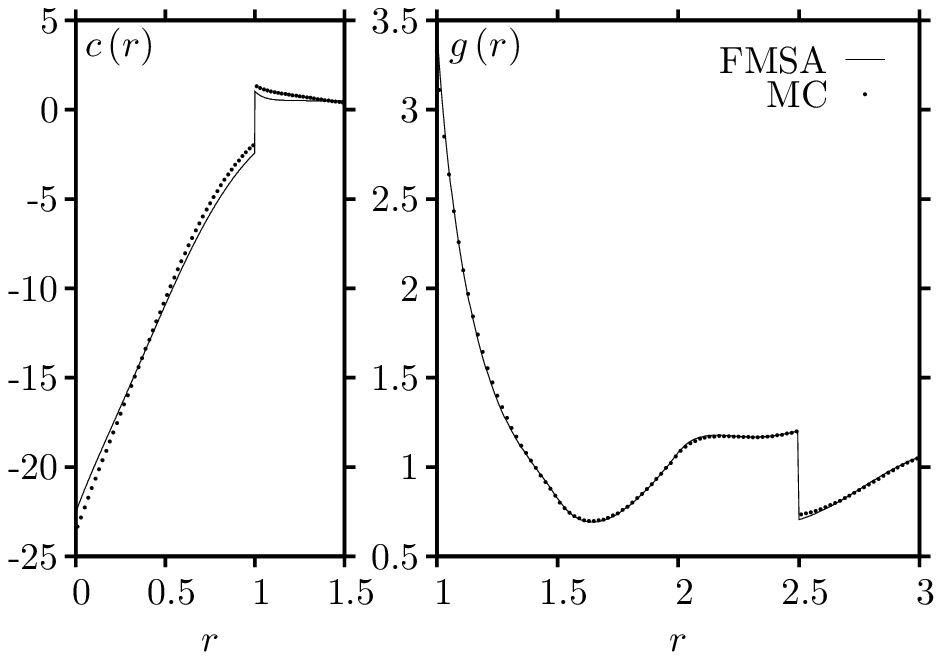}
\end{center}
\caption{}
\label{fig:dcf_rdf_l25}
\end{figure}

\textbf{\noindent Fig.~\ref{fig:dcf_rdf_l25} Full DCF (left panel) and full
RDF (right panel) for SW fluid with $\lambda=2.5$, $\varepsilon\beta=0.5$
and $\rho=0.75$. Lines denote FMSA results, symbols denote simulation
results.}

\begin{figure}[tbp]
\begin{center}
\includegraphics[width=12cm, clip]{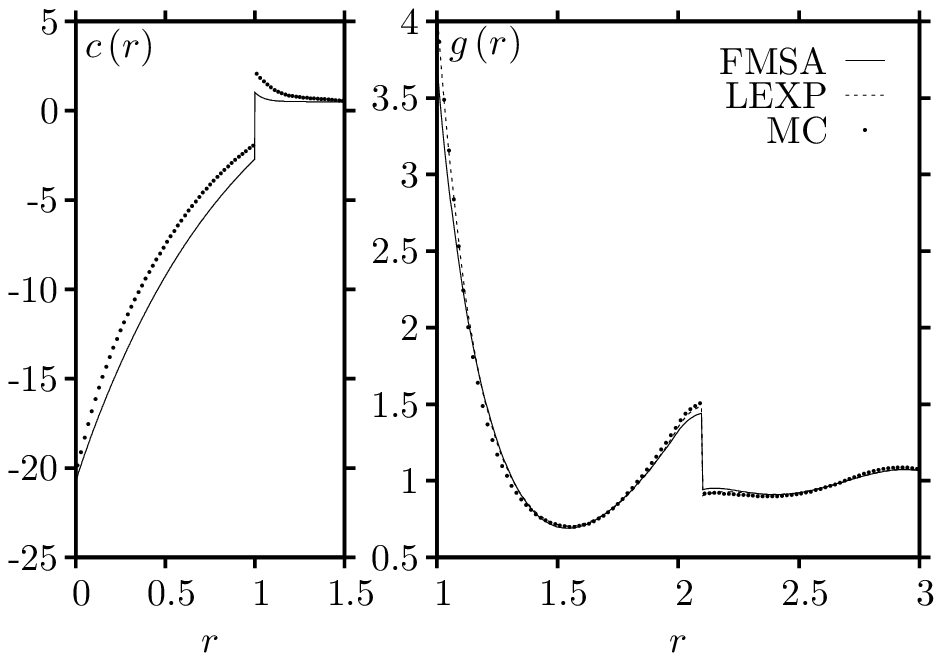}
\end{center}
\caption{}
\label{fig:dcf_rdf_l21}
\end{figure}

\textbf{\noindent Fig.~\ref{fig:dcf_rdf_l21} Full DCF (left panel) and full
RDF (right panel) for SW fluid with $\lambda=2.1$, $\varepsilon\beta=0.5$
and $\rho=0.75$. Lines denote FMSA results, symbols denote simulation
results. Dotted line denotes linearized exponential approximation (LEXP).}

\begin{figure}[tbp]
\caption{}
\label{fig:Compress}
\includegraphics[width=12cm, clip]{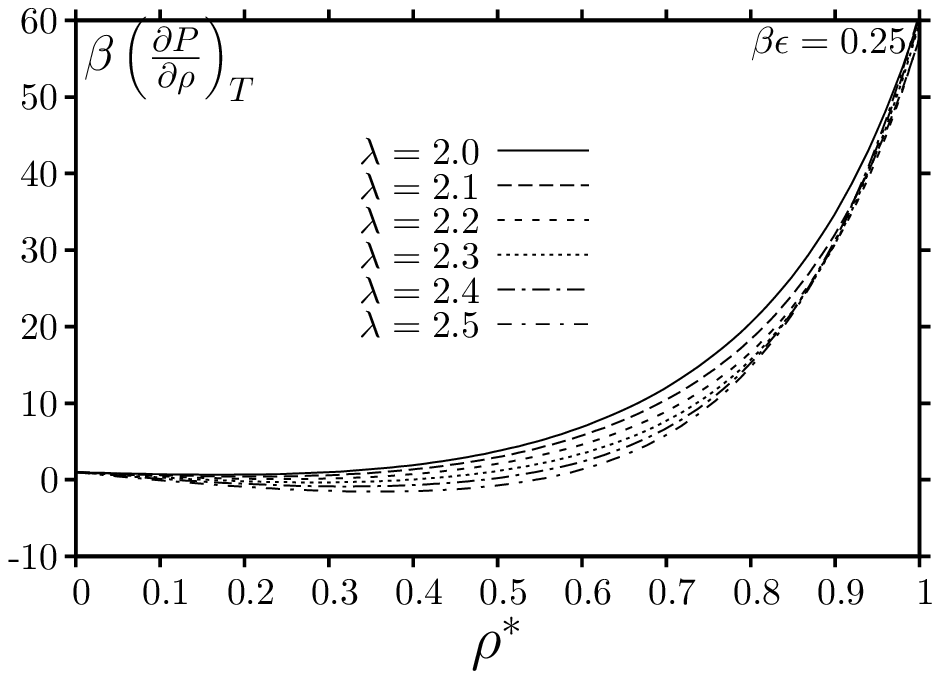} 
\includegraphics[width=12cm,
clip]{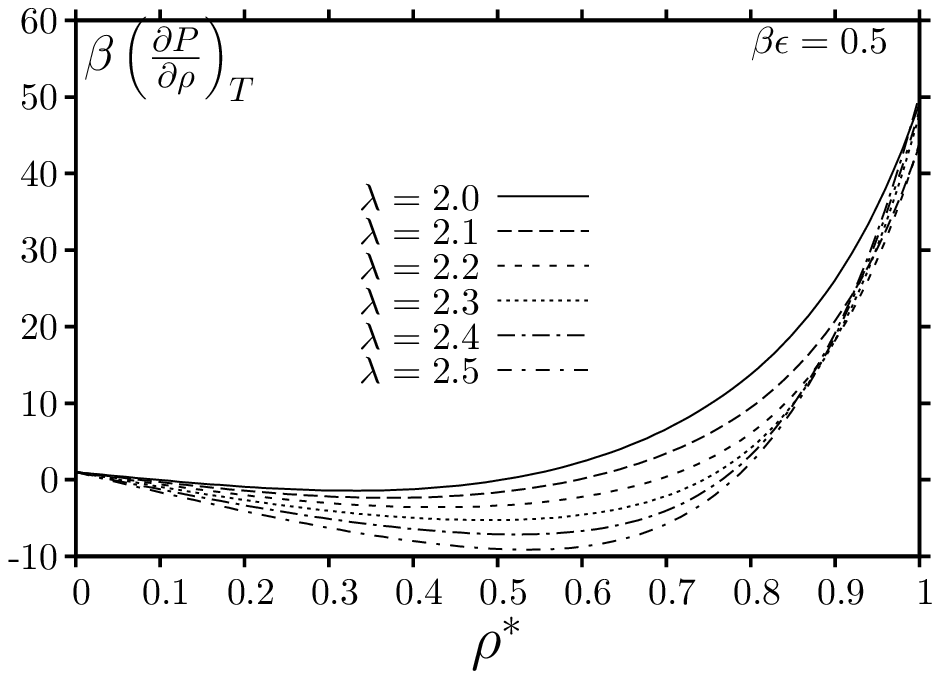} 
\end{figure}

\textbf{\noindent Fig.~\ref{fig:Compress} Inverse reduced isothermal
compressibility $\chi_T^{-1}=\beta\left( \frac{\partial P}{\partial\rho}
\right)_{T}$ of the SW fluids at $\varepsilon\beta=0.5$ and $%
\varepsilon\beta=0.25$ .}

\begin{figure}[tbp]
	\begin{center}
		\includegraphics[width=12cm, clip]{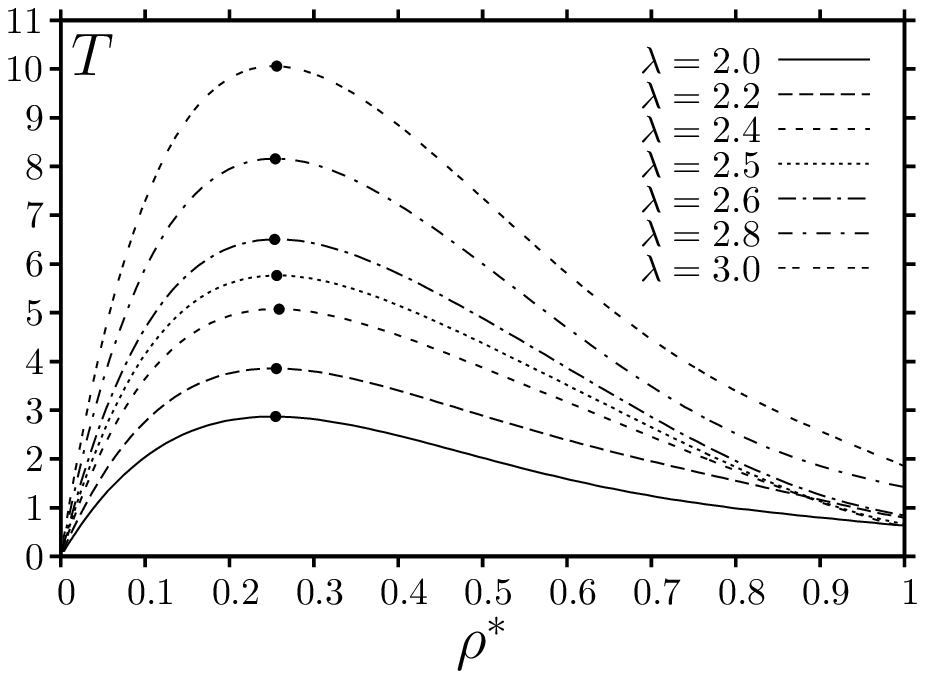}
	\end{center}
	\caption{}
	\label{fig:spinodal}
\end{figure}

\textbf{\noindent Fig.~\ref{fig:spinodal} Spinodal curves for SW fluids
with $\lambda=2.0, 2.2, 2.4, 2.5, 2.6, 2.8, 3.0$.}

\newpage

\newpage

\newpage

\newpage

\newpage

\end{document}